\documentclass[twoside,twocolumn,9pt]{article}
\usepackage{extsizes}
\usepackage[version=3]{mhchem}
\usepackage[left=1.5cm, right=1.5cm, top=1.785cm, bottom=2.0cm]{geometry}
\usepackage{flushend}
\usepackage{times,mathptmx}
\usepackage{sectsty}
\usepackage{lastpage}
\usepackage[format=plain,justification=justified,singlelinecheck=false,font={stretch=1.125,small,sf},labelfont=bf,labelsep=space]{caption}
\usepackage{float}
\usepackage{fancyhdr}
\usepackage{fnpos}
\usepackage[english]{babel}
\addto{\captionsenglish}{%
	
}
\usepackage{array}
\usepackage{droidsans}
\usepackage{charter}
\usepackage[T1]{fontenc}
\usepackage[usenames,dvipsnames]{xcolor}
\usepackage{setspace}
\usepackage[compact]{titlesec}
\usepackage{hyperref}
\usepackage[latin1]{inputenc}  
\usepackage{rsc}  
\usepackage{multicol} 
\usepackage{mwe} 
\usepackage{rotating}  
\usepackage{graphicx,subfigure}  
\usepackage{graphicx}  
\usepackage{placeins} 

\usepackage{tikz}				
\usetikzlibrary{topaths,arrows}
\usetikzlibrary[topaths]

\usepackage{tabstackengine}
\usepackage{cuted}

\definecolor{cream}{RGB}{222,217,201}

\begin{document}
\pagestyle{fancy}   
\thispagestyle{plain}
	
	
\makeFNbottom
\makeatletter
\renewcommand\LARGE{\@setfontsize\LARGE{15pt}{17}}
\renewcommand\Large{\@setfontsize\Large{12pt}{14}}
\renewcommand\large{\@setfontsize\large{10pt}{12}}
\renewcommand\footnotesize{\@setfontsize\footnotesize{7pt}{10}}
\makeatother
	
\renewcommand{\thefootnote}{\fnsymbol{footnote}}
\renewcommand\footnoterule{\vspace*{1pt}%
\color{cream}\hrule width 3.5in height 0.4pt \color{black}\vspace*{5pt}} 
\setcounter{secnumdepth}{5}
	
\makeatletter 
\renewcommand\@biblabel[1]{#1}            
\renewcommand\@makefntext[1]%
{\noindent\makebox[0pt][r]{\@thefnmark\,}#1}
\makeatother 
\renewcommand{\figurename}{\small{Fig.}~}
\sectionfont{\sffamily\Large}
\subsectionfont{\normalsize}
\subsubsectionfont{\bf}
\setstretch{1.125} 
\setlength{\skip\footins}{0.8cm}
\setlength{\footnotesep}{0.25cm}
\setlength{\jot}{10pt}
\titlespacing*{\section}{0pt}{4pt}{4pt}
\titlespacing*{\subsection}{0pt}{15pt}{1pt}
	
	
\makeatletter 
\newlength{\figrulesep} 
\setlength{\figrulesep}{0.5\textfloatsep} 
	
\newcommand{\topfigrule}{\vspace*{-1pt}%
\noindent{\color{cream}\rule[-\figrulesep]{\columnwidth}{1.5pt}} }
	
\newcommand{\botfigrule}{\vspace*{-2pt}%
\noindent{\color{cream}\rule[\figrulesep]{\columnwidth}{1.5pt}} }
	
\newcommand{\dblfigrule}{\vspace*{-1pt}%
\noindent{\color{cream}\rule[-\figrulesep]{\textwidth}{1.5pt}} }
	
\makeatother
	
\twocolumn [
\begin{@twocolumnfalse}
\vspace{3cm}
\sffamily
\noindent\LARGE{\textbf{The Role of Energy Cost on Accuracy, Sensitivity, Specificity, Speed and Adaptation of T Cell Foreign and Self Recognition}} \\ 
\vspace{0.3cm}  \vspace{0.3cm} \\ 
		
\noindent\large{Gyubaek Shin\textit{$^{a}$} and Jin Wang\textit{$^{a}$}}\textit{$^{b}$}\textit{$^{\ast}$} \\

\noindent\normalsize{\indent The critical role of energy consumption in biological systems including T cell discrimination process has been investigated in various ways. The kinetic proofreading(KPR) in T cell recognition involving different levels of energy dissipation influences functional outcomes such as error rates and specificity. In this work, we study quantitatively how the energy cost influences error fractions, sensitivity, specificity, kinetic speed in terms of Mean First Passage Time(MFPT) and adaption errors. These provide the background to adequately understand T cell dynamics. It is found that energy plays a central role in the system that aims to achieve minimum error fractions and maximum sensitivity and specificity with the fastest speed under our kinetic scheme for which numerical values of kinetic parameters are specially chosen, but such a condition can be broken with varying data. Starting with the application of steady state approximation(SSA) to the evaluation of the concentration of each complex produced associated with KPR, which is used to quantify various observables, we present both analytical and numerical results in detail.} \\
		
		
\end{@twocolumnfalse} \vspace{0.6cm}
]
	
\renewcommand*\rmdefault{bch}\normalfont\upshape
\rmfamily
\section*{}
\vspace{-1cm}

	
\footnotetext{\textit{$^{a}$~Department of Chemistry, SUNY Stony Brook - 100 Nicolls Road, Stony Brook, NY 11794, USA}}
\footnotetext{\textit{$^{b}$~Department of Physics and Astronomy, SUNY Stony Brook - 100  Nicolls Road, Stony Brook, NY 11794, USA. E-mail: jin.wang.1.@stonybrook.edu}}

	

	

\section{Introduction}
\indent One of the well known biological malfunctions is the deviation from the normal condition of being able to maintain the ability to efficiently differentiate foreign antigens from self-proteins attacking living cells. It may be associated with an abnormality of KPR processes, which prevents a bound form of ``wrong'' ligands from being dissociated at a sufficiently high rate\cite{1}. The affinity ratio of ``correct'' and ``wrong'' ligands with T cell receptor is typically a measurable quantity that determines the efficiency of such dissociation. Hopfield and Nino developed KPR theory in biosynthetic processes\cite{2,3}. Hopfield formulated error fractions for protein synthesis\cite{2}. They elucidated that enzymes discriminate two different reaction pathways, leading to correct or incorrect products due to KPR. Since then, extensive researches on sensitivity and specificity associated with error fractions have been performed. Goldbeter et al found that covalent modification in protein involving biological systems affects a sensitivity amplification using Steady State Approximation(SSA)\cite{4}.\\
\indent A series of modifications after ligand binding in the KPR process involves extra steps which creates ``time delay'' $\tau$. The extra steps leading to signaling are critical factors that allow for reduction in error rates, indicating high efficiency of kinetic discrimination\cite{5}.  However, KPR also involves free energy cost for activation of an initially formed complex, which occurs in nonequilibrium states\cite{1}. 
The energy is also crucial in reducing the error rates and allowing increased specificity\cite{2}. Before KPR attracted great interests, there had been several studies focusing on the effect of energy cost for KPR in biological processes such as tRNA aminoacylation\cite{6,7} and so on. \\
\indent Beyond the classical studies on discrimination process for biological systems such as Hopefiled\cite{2}, Nino\cite{3}, and McKeithan\cite{8}, there has been a fair amount of accomplishment on T cell recognition with certain modifications \cite{9,10,11}, which make it possible to address several deficiencies found in existing models. For example, Qian calculated an error fraction depending on both KPR steps and energy cost using the Successive Rapid Equilibrium Approximation (SREA) by assuming that there is energy input for only the first cycle.\cite{9}. Chen et al. are the ones provided a the formulation of T cell sensitivity and specificity in a quantitative manner, depending on the number of KPR steps using a SSA\cite{12}. There have been significant contributions from Cui\cite{10} and Banerjee\cite{11}, focusing on the detailed relationship between error rates and MFPTs. Despite their efforts on detailed analysis of the dynamics, their studies are based on the kinetic scheme in terms of only energy cost, lacking comprehensive information for which both KPR steps and energy dissipation are taken into consideration. For convenience, we use the term ``KPR steps'' instead of phosphoylation steps although technically a KPR process includes both phosphorylations and the dissociation of each intermediate product.
Here, the following questions can be raised:\\
(1) If energy consumption plays an central role in reducing errors, how does energy influence sensitivity and specificity in T cell discrimination process, and what are the relationships among error fraction, sensitivity and specificity under identical conditions? \\
(2) Although two factors, the KPR steps and energy dissipation, both of which contribute to editing process of the system have
opposite effects in terms of the time required to complete the associated process, the retardation due to the increased KPR steps may be mitigated by sufficient level of energy dissipation. Can the MFPT data provide adequate information to determine such energy level under the given condition? \\
(3) Against the initial stimulus signal, can a given T cell dynamics achieve a virtually full recovery which may be characterized by measuring output? What is the role of energy cost in such a process? \\
\indent In order to answer the above questions, we design a kinetic model describing T cell discrimination process. After introducing the chosen model for our discussion, this paper shows a detailed procedure leading to analytical expressions for these quantities based on the SSA by imposing energy input in ``every step'' since a series of modifications that occur after ligand binding requires energy consumption and is out of equilibrium\cite{1}. Based on the kinetic model, we calculated error fractions, sensitivity and specificity in terms of both KPR steps and energy cost. \\
\indent We also calculated the kinetic speed in terms of MFPT, the average time required to complete the signaling event starting from an initial state, depending on energy with given KPR steps. The entire picture of the dynamics in T cell recognition will still remain unclear with the only sensitivity and specificity data available until the consequences of MFPT are evaluated. This is because the MFPT provides information on the time required for signaling to be completed under energy dissipation. We also see how energy influences the adaptation errors in response to the shift of a particular parameter, which is the rate constant used in our kinetic model. \\

\vspace{5mm}

\section{Kinetic Scheme describing T Cell Recognition}

\indent The following detailed kinetic scheme reflects KPR associated with the energy consumption. This scheme is based on the McKeithan's kinetic model.\cite{8,9}, but we provide a modified version of the scheme by incorporating two additional paths such as the backward reactions between intermediate complexes and the direct formation steps\cite{2,10} in the real context of the biological system.	

\begin{figure*}[!]
	\centering
	\subfigure[\normalsize{\textbf{\textcolor{blue}{Foreign recognition of T cell}}}]{
		\includegraphics[width=0.67\textwidth]{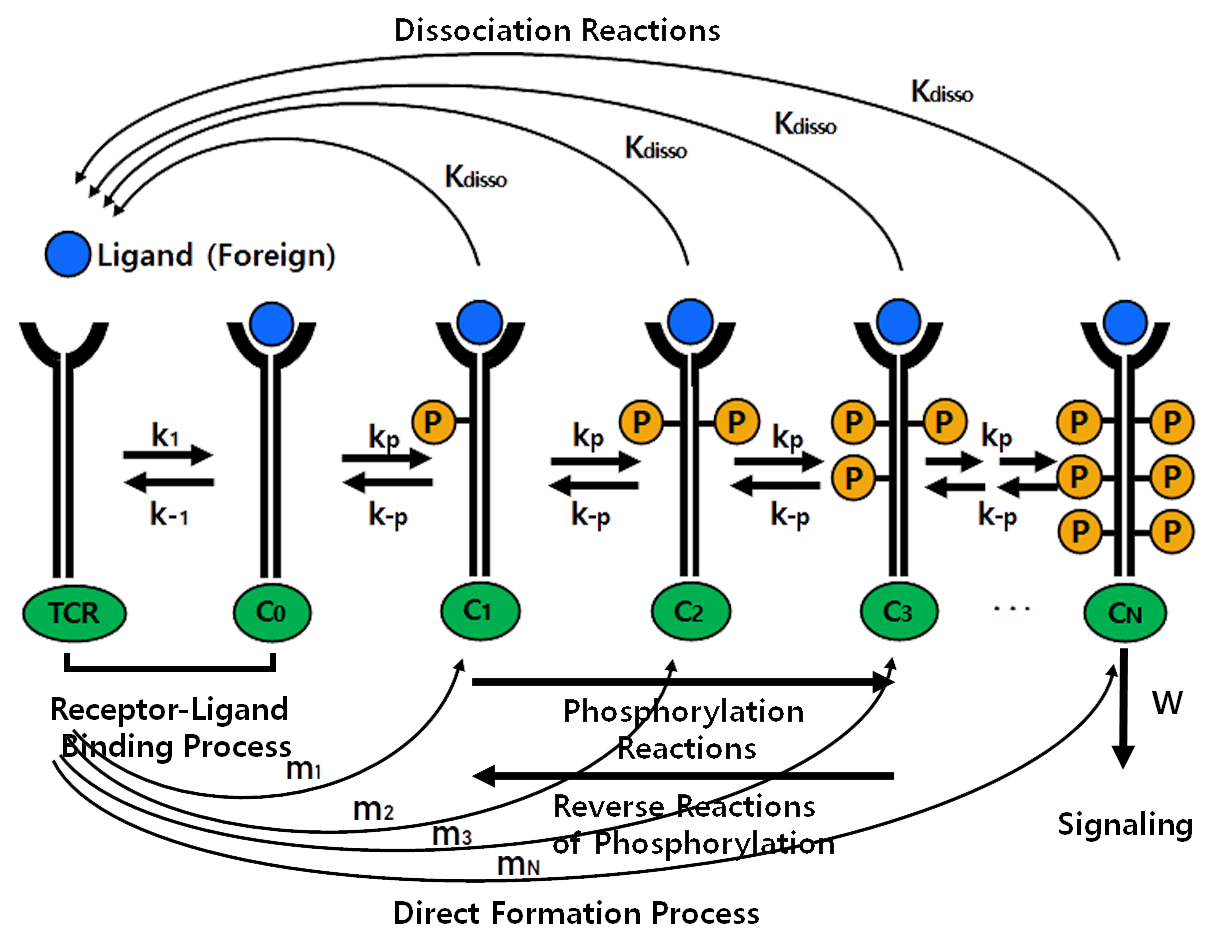} 
	}  \\ 
	\subfigure[\normalsize{\textbf{\textcolor{red}{Self recognition of T cell}}}]{
		\includegraphics[width=0.67\textwidth]{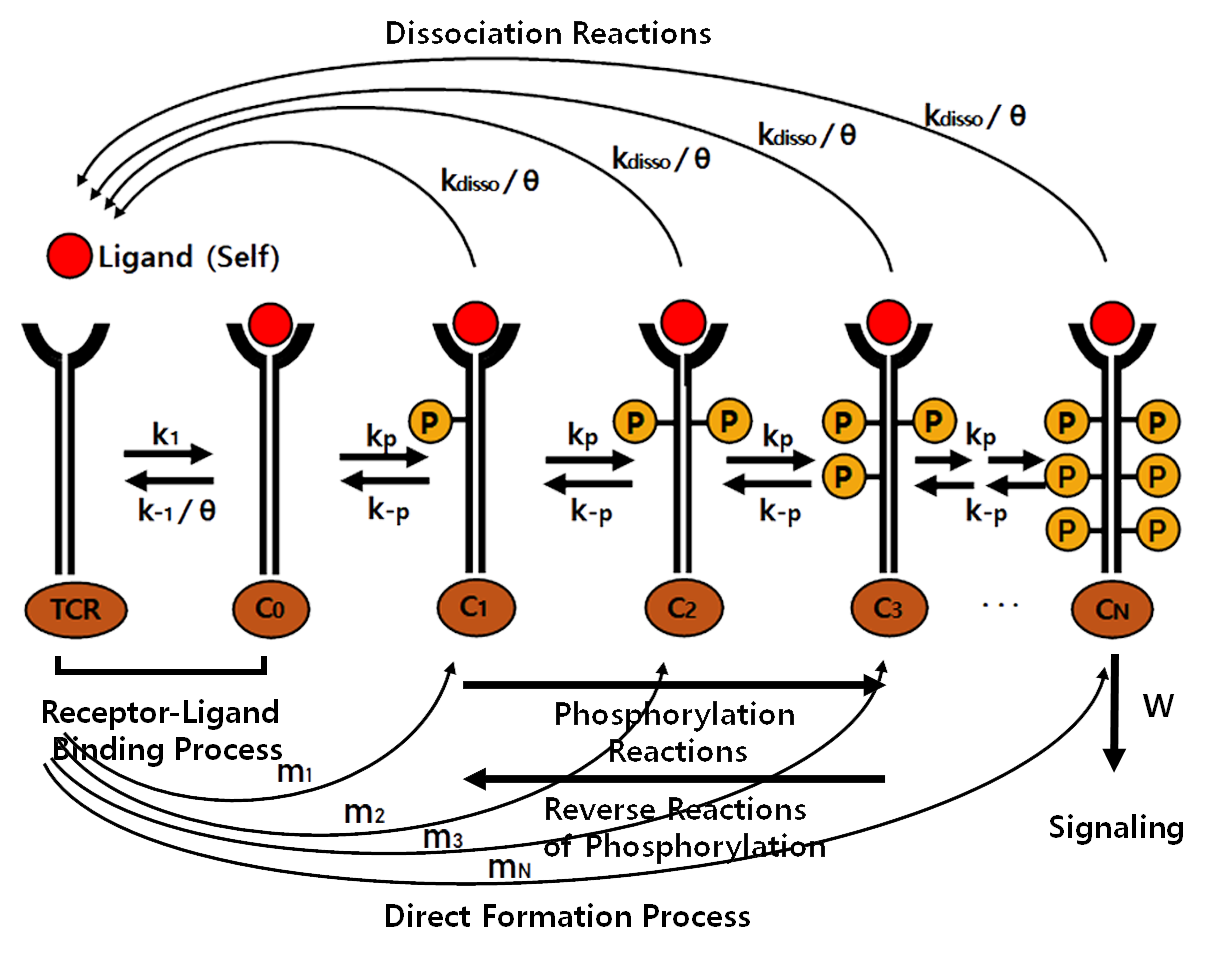}
	}
	\caption{\textit{\textbf{Schematics of the kinetic model for ``N'' kinetic proofreading process (a): foreign ligands  (b) : self ligands} \quad Followed by receptor-ligand binding, a series of modifications triggers the T cell recognition signaling. At each intermediate state, self ligands dissociate $1/\theta$ ($\theta$=0.1 in our case) times faster than foreign ligands. The $k_{1}$ and $k_{-1}$ are both forward and backward rate constants at fast equilibrium. The governing rate constants between two intermediate complexes after the receptor-ligands binding step are $k_{p}$ and $k_{-p}$ respectively. Each $m_{i}$ denotes the rate of the direct formation starting from free ligands. Coupling to energy source is considered in every step.}}			
\end{figure*}
	
\indent The initial complex formed by a T cell receptor and equal amounts of foreign and self ligands triggers a series of modifications, leading to signaling. Since the first complex reaches equilibrium rapidly, the values of governing rate constants $k_{1}$ and $k_{-1}$ for the corresponding forward and backward reactions, respectively are substantially higher than the ones given by $k_{p}$ and $k_{-p}$ for the rest of the reactions. The dissociation events at each intermediate complex leading to its initial state with the rate of $k_{-i}^{*}$ (i=1,2,...N) allow for a reduction in the amount of ligand bound molecules. We set the same value of the dissociation constant($k_{disso}$) for each intermediate complex for simplicity. There is a need to incorporate the rate constants governing the direct formation process, which leads to the development of the complexes without passing through earlier steps into the full ``rate equation''. The direct formation constants denoted by $m_{i}(i=1,2,...N)$ decrease with the KPR steps due to the higher energy intermediates as KPR progresses\cite{2}. In addition to this, they also decrease with consumed energy according to the formula for energy dissipation. We allow variation of the backward rates and the direct formation rates such that they decrease with energy. However, the reverse rate constant $k_{-1}$ that is associated with the fast equilibrium is unchanged. The transfer rate ``W'' is included in the irreversible process from the final complex to the absorbing site where the associated dynamics is completed. The equilibrium ATP and ADP concentrations are related to the rate constants\cite{9,13,14,15}. 
\begin{equation}
\frac{[ATP]_{eq}}{[ADP]_{eq}}=\frac{k_{-p}^{\circ}[C_{1}]_{eq}}{k_{p}^{\circ}[C_{0}]_{eq}}=\frac{k_{-p}^{\circ}m_{1}k_{-1}}{k_{p}^{\circ}k_{1}k_{-1}^{*}}
\end{equation}
, where $k_{p}$ and $k_{-p}$ are pseudo first order rate constants denoted by $k_{p}=k_{p}^{\circ}[ATP]$ and $k_{-p}=k_{-p}^{\circ}[ADP]$ respectively since we assume both ATP and ADP are in excess. In addition to this, we impose the condition that the amount of free ligands is much greater than free receptors such that the reaction starting from the free receptor-ligands state follows the pseudo first order kinetics. \\
\indent The second equality comes from the relationship between the ratio of the equilibrium concentration and kinetic constants (i.e) $\frac{[C_{1}]_{eq}}{[C_{0}]_{eq}}=\frac{k_{-1}m_{1}}{k_{-1}^{*}k_{1}}$.
	
The free energy of ATP hydrolysis is given as
\begin{equation} 
\Delta G_{DT}=\Delta G_{DT}^{\circ}+RTln\left(\frac{[ATP]}{[ADP]}\right)=RTln\left(\frac{k_{p}k_{1}k_{-1}^{*}}{k_{-p}m_{1}k_{-1}}\right)=RTln\gamma
\end{equation}
	
, where 
	
\begin{equation}
\Delta G_{DT}^{\circ}=-RTln \left(\frac{[ATP]_{eq}}{[ADP]_{eq}}\right)
\end{equation}
	
\noindent and we can define $\gamma$ as the available dimensionless free energy from each ATP hydrolysis as\cite{9,13,14}.  
	
\begin{align}
\gamma &= \frac{k_{p}k_{1}k_{-1}^{*}}{k_{-p}m_{1}k_{-1}}  \hspace{1.8em} (N=1) \\
\gamma &= \frac{k_{p}k_{-(i+1)}^{*}m_{i}}{k_{-p}m_{i+1}k_{-i}^{*}}  \hspace{1.0em} (N>1, i+1=N) 
\end{align}
, where $k_{-i}^{*}$(i=1,2,...N) is all the same. 
	
Although the energy $\gamma$ is associated with many kinetic variables in principle, we explore two kinetic parameters $m_{i}$ and $k_{-p}$ related to the energy $\gamma$.
	
In other words,
\begin{align}
m_{1} &= \frac{k_{p}k_{1}k_{disso}}{\gamma k_{-p}k_{-1}}  \hspace{2.5em} (N=1) \\
m_{i+1} & \hspace{0.75em} = \frac{k_{p}m_{i}}{\gamma k_{-p}} \hspace{4.0em} (N>1, i+1=N) 
\end{align}
	
In addition to this, we allow the backward rate constants $k_{-p}$ that decrease with energy input $\gamma$ in the following fashion. The dimensionless free energy $\gamma$=1 indicates that the system is governed by the detailed balance condition but as the system becomes out of equilibrium, the backward rates get less dominant. Such a kinetic scheme makes it possible to the direct formation constant $m_{i}$ to decrease with both KPR steps and the energy.

\begin{table}[h]
	\begin{center}
		\begin{tabular}{|c|c|}
			\hline
			$\gamma$ & \multicolumn{1}{|c|}{Relative Values of $k_{-p}$} \\
			\hline
			1 & \multicolumn{1}{|c|}{1.1} \\
			\hline
			10 & \multicolumn{1}{|c|}{1/5} \\
			\hline
			$10^{2}$ & \multicolumn{1}{|c|}{1/10} \\
			\hline
			$10^{n}$ (n$\geq$3) & \multicolumn{1}{|c|}{1/(50 x $5^{n-3}$)}  \\	
			\hline		
		\end{tabular}		
	\end{center}
\end{table}

The affinity ratio between ``wrong''(self-protein) and ``correct''(foreign antigen) for targeting is given by
	
\begin{equation}
\theta=\frac{\frac{C_{i}^{'}}{[R][L]}}{\frac{C_{i}}{[R][L]}}
=\frac{\frac{m_{i}^{'}}{k_{disso}^{'}}}{\frac{m_{i}}{k_{disso}}}   
=\frac{k_{disso}}{k_{disso}^{'}}   
=\frac{k_{-1}}{k_{-1}^{'}}
\end{equation}
assuming $m_{i}$ is the same for both the correct and wrong ligands. The self-proteins bound to the receptor dissociate more quickly than the foreign antigens indicating that the affinity ratio is less than 1. We set the value of $\theta$ to be 0.1 which is consistent with experimental observation.\cite{1} The numerical values of kinetic parameters we take are as follows: R=10, L=10$^{3}$, $k_{1}^{\circ}$=10$^{4} s^{-1}$, $k_{-1}$=9x10$^{3} s^{-1}$, $k_{p}$=0.5$s^{-1}$, $k_{disso}$=0.1$s^{-1}$ and W=$10^{-5}s^{-1}$. Experimental evidence indicates that the typical range for binding rate is from O(10$^{2}$) to O(10$^{6}$)$s^{-1}$ and for unbinding rate is around O(10$^{-1}$)$s^{-1}$ \cite{16}. \\

\vspace{5mm}

\section{Result : Error Fractions}
	
The Nonequlibrium Steady States(NESS) are typically sustained by constant sources and flux. The system we design is in NESS that are maintained with constant concentration of ATP, ADP and the free ligands. Such a system differs from the one subject to the condition for which the number of molecules fluctuates\cite{13,15}. The detailed procedure that leads to error fractions is given in Appendix A. \par
	
\begin{figure}[!] 
	\subfigure[]{
		\includegraphics[width=0.5\textwidth]{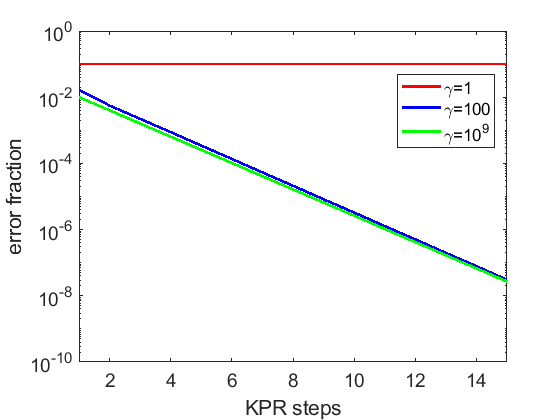}
	}
	\quad
	\subfigure[]{
		\includegraphics[width=0.5\textwidth]{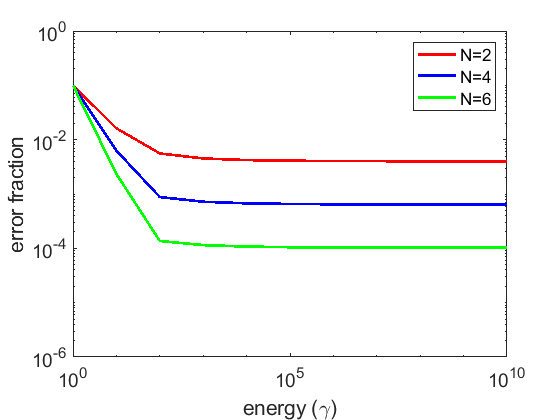}			
	}
	\caption{\textit{(a) The red, blue and green lines are the error rates depending on KPR steps for the energy $\gamma$= 1,100, and 10$^9$ respectively. When there is no energy input, the error rate does not change with the number of phosphorylated products. (b) The red, blue and green lines are the error rates for the KPR steps of 2,4,and 6 respectively with energy $\gamma$, featuring their continuous drops until energy reaches around 100.}}	
\end{figure}
	
\indent We obtained the numerical results for error fractions depending on KPR steps and energy consumption, featuring their decrease with both factors(See Figure 2(a) and (b)). The error fractions decline drastically until energy $\gamma$ reaches around 100, giving a disparity between each KPR step. However, after this point, they quickly converge their minimum values. T cell reduces error rates by recognizing foreign antigens with the help of multiple phosphorylation steps and certain amount of energy expenditure even though the misrecognition of self-proteins as foreign peptides commonly occurs. We find that lower error rates can be achieved when the forward rate or the affinity ratio decreases. The reduced forward rate allows the T cell system to regulate the formation of the self ligand complexes sufficiently compared to the foreign ligand products, yielding the lower error rates. This indicates the system has more time to fix errors by differentiating self proteins from foreign antigens. Such a recognition process can be facilitated with the decline in the affinity ratio, which is associated with the enhanced specificity. We find when the forward rate constant $k_{p}$ changes from 0.5 to 0.1, the error rate is reduced by 31.0\% for N=2 and 64.0\% for N=6, both of which are subject to the energy cost $\gamma$ being 1000. On the other hand, when the affinity ratio is reduced by 50\%, the corresponding error rates become 7.62x$10^{-4}$ and 2.55x$10^{-6}$ respectively for the two different KPR cases with the same energy input, which implies its decreases by 83.1\% and 99.8\% for each case. Experimentally, it is well known that the typical error fraction is less than $10^{-6}$ under the condition when the affinity ratio $\theta$ is 0.1 which is typical in human body\cite{1}. Another study estimating the error rate based on a simple kinetic proofreading model suggests that the rate is approximately $10^{-4}$ at the affinity ratio of 0.01 for N=4\cite{8}. No information on the number of phosphorylation steps is available for both cases.
	
\vspace{5mm}

\section{Result : Sensitivity and Specificity} 
	
\indent Both sensitivity and specificity based on the kinetic model were computed. Sensitivity is defined as the probability of having the number of foreign antigens sufficient to generate major signaling out of the total complex. On the other hand, specificity is defined as a factor to determine the ability to discriminate the correct ligands (foreign antigens) from the wrong ones (self-proteins) in their active states which contribute to major signaling\cite{12}.\\ Chan et al. provided a simple expression for these quantities in kinetic proofreading in the context of T cell recognition in the following manner\cite{12}.  We directly follow the procedures they present.
	
This implies the definition of sensitivity and specificity can be expressed as follows: \\
Sensitivity=TP/(TP+FN) \\
Specificity=TP/(TP+FP) \\
where \\
TP = The number of signaling events for a ``correct'' ligand \\
FN = The number of zero signaling events for a ``correct'' ligand \\
TN = The number of zero signaling events for a ``wrong'' ligand \\
FP = The number of signaling events for a ``wrong'' ligand \\
	
If we simply use the fraction of the active complexes, taking $\frac{C_{N}}{C_{total}}$ as $\alpha^{N}$, then \\
TP= $C_{total,foreign}\;\alpha_{correct}^{N}$ \\
FP = $C_{total,self}\;\alpha_{wrong}^{N}$\\
FN= $C_{total,foreign}\;\left(1-\alpha_{correct}^{N}\right)$ \\
TN = $C_{total,self}\;\left(1-\alpha_{wrong}^{N}\right)$ \\
	
Therefore, \\
	
\begin{equation}
\textnormal{Sensitivity}=\frac{C_{total,foreign}\;\alpha_{correct}^{N}}{C_{total,foreign}\;\alpha_{correct}^{N}+C_{total,foreign}\;\left(1-\alpha_{correct}^{N}\right)}
\end{equation}
	
\begin{equation}
\textnormal{Sepcificity}=\frac{C_{total,foreign}\;\alpha_{correct}^{N}}{C_{total,foreign}\;\alpha_{correct}^{N}+C_{total,self}\;\alpha_{wrong}^{N}}
\end{equation}
	
Here, $C_{total}$ can be achieved by adding the concentrations of all intermediates including the ligand-receptor complex at the final state, which is taken from both foreign and self-ligands, sorted by different ``N''. The associated concentrations of foreign ligands for the purpose of numerical calculation were taken from the equation (9) to (17). Qualitative speaking, sensitivity is the percentage of the amount of the ``true positive'' products (the active foreign ligand complexes that contribute to signaling) out of the total products which also include ``false negative'' products (foreign complexes that do not respond). On the other hand, specificity is the percentage of the amount of the ``true positive'' products out of the total products responsible for signaling, which are generated from both foreign and self proteins. \par
\indent Chan and et al.\cite{12} shows the feature of decrease in sensitivity depending on the number of KPR steps based on their idealized kinetic scheme for which reverse reactions between intermediate states are not taken into account without using energy $\gamma$. They also obtained the result through increased specificity, reaching to 1.0 depending on the number of KPR steps. The trade-off between sensitivity and specificity is also observed in our model\cite{12,17}. 
	
\begin{figure*}[!] 
	\subfigure[]{
		\includegraphics[width=0.5\textwidth]{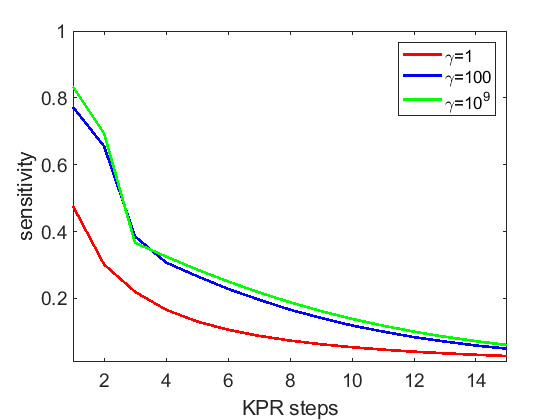}
	}
	\quad
	\subfigure[]{
		\includegraphics[width=0.5\textwidth]{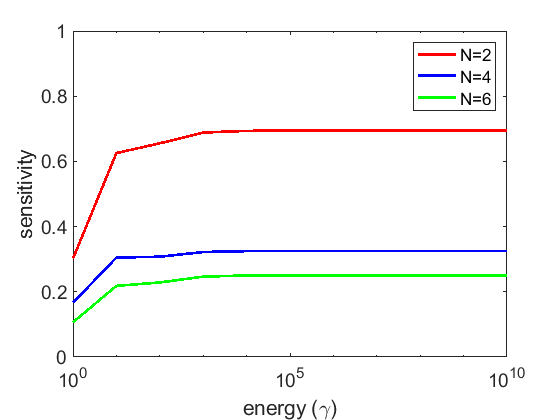}			
	}
	\caption{\textit{(a) The sensitivity decreases with KPR steps, but applied energy is responsible for higher sensitivity. (b) The sensitivities for different number of KPR steps with energy $\gamma$ reveals their initial growth with energy input.}}
\end{figure*}

\begin{figure*}[!] 
	\subfigure[]{
		\includegraphics[width=0.5\textwidth]{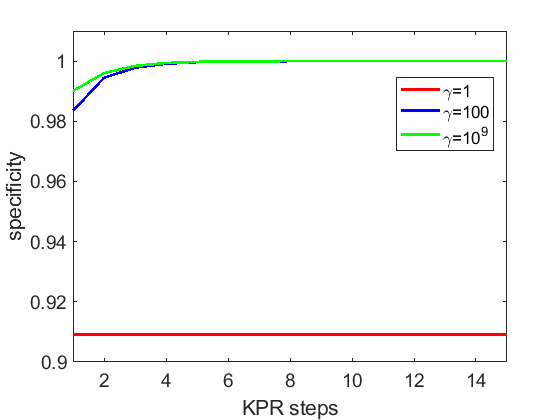}
	}
	\quad
	\subfigure[]{
		\includegraphics[width=0.5\textwidth]{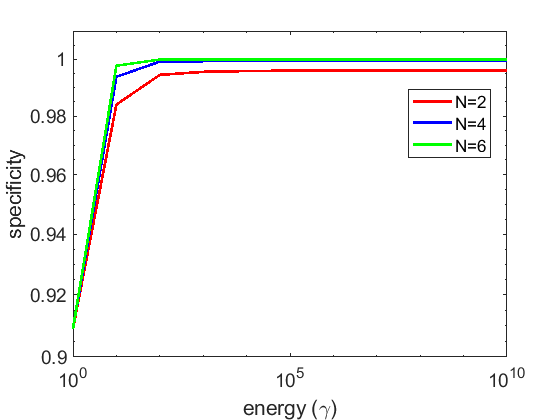}			
	}
	\caption{\textit{(a) The growth of specificity with KPR steps at different energy $\gamma$. When the system is governed by the detailed balance condition($\gamma$=1), the specificity does not vary with the number of phosphorylated products. (b) The specificities for different number of KPR steps with energy. They approach maximum values quickly with energy supply.}}		
\end{figure*}
	
\indent As shown in Figure 3(b), our results show that the sensitivity increases and converges to a certain value as the energy is consumed for a given number of KPR steps. Such behavior of the sensitivity in terms of the energy input can be interpreted as follows: \\
(1) Its initial growth is associated with the immediate drop of the backward rate which is responsible for the increase in the concentration of each complex due to the energy $\gamma$. \\
(2) However, its relative robustness after the point is because of the balance between decreases in both the backward and the direct formation rates which have the opposite effects in terms of the change in the concentration of products as more energy is supplied.
As indicated in the general expression for sensitivity, the growth in the amount of final complex relative to the concentration of total products increases the sensitivity. \par	
It is also found that the sensitivity decreases with the number of KPR steps for a given energy(See Figure 3(a)), which is generally consistent with the results from Chan and et al.\cite{12}. It is noticeable that under the detailed balance condition($\gamma$=1), the sensitivity has a low value even when a small number of phosphorylated products are formed. The specificity obtained from our model using SSA has the feature approaching a maximum value rapidly as the energy cost $\gamma$ increases(See Figure 4(b)). We observe that the number of KPR steps does not affect the specificity in a significant manner with given energy $\gamma$, showing marginal growth of the quantity as N increases(See Figure 4(a)). We find the rapid increase of specificity converging to the approximate value of 1.0 with energy. However, it is remarkable when there is no energy input, the specificity does not change with KPR steps. \par
\indent Based on our kinetic scheme, the sensitivity for N=2 at the energy $\gamma$=1000 are 0.689 which is in good agreement with the value estimated from the Mckeithan's case which is 0.694. However, when multiple phosphorylation is involved, we find a slight deviation. For example, N=6, we get 0.247 of the sensitivity which is lower than the other result which reads 0.335. On the other hand, the specificity in our model estimated at $\gamma$=1000 for N=2 and N=6 are 0.996 and 0.999 respectively which are slightly higher than the corresponding values from the other model subject to the identical values of parameters, which read 0.862 and 0.996 respectively\cite{8}. The direct comparison between the two models is not valid because the formalism based on Mckeithan's case does not contain the parameter for energy cost. When the forward rate constant doubles which is $k_{p}$ of 1.0, we observe the enhanced sensitivity. To be specific, when the energy $\gamma$=1000, the corresponding sensitivities are 0.821 and 0.357 for N=2 and N=6 respectively, which reveals the increase of the sensitivity by at least 40\% compared to the output with the original value of $k_{p}$ which are 0.689 and 0.247 respectively. We find that such an increase is pronounced when large number of phosphorylation products are involved. The increase in the forward rate is responsible for the growth in the formation of each phosphorylated product, yielding higher sensitivity. For specificity, when we take $\theta$=0.3 instead of $\theta$=0.1, we find the decease of the specificity by around 6\% and 2\% respectively for N=2 and N=6. As indicated from the results, the loss of sensitivity is compromised by growth of specificity as more phosphorylated products are formed. The trade-off between the two physical outcomes are already discussed in Chan et al\cite{12}.

\vspace{5mm}

\section{Result: Mean First Passage Time}

\indent The speed of KPR cascade associated with Mean First Passage Time(MFPT) provides information on how rapidly the immune system responds to the foreign ligand. More precisely speaking, it is the average time taken to produce the final product that contributes immediate signaling from foreign antigens\cite{10}. We find that the energy input and the number of KPR steps are the major factors that determine the MFPT in KPR model. Our work in this part is directly towards the evaluation of the passage time depending on the energy consumption when the foreign ligands are involved in KPR. Similar works have been done by Banerjee et al. for the calculation of MFPT of DNA replication process\cite{11}. However, it is based on a different style of biological network that takes separate mechanisms relying on the type of ligands, both correct and incorrect ones forming associated complexes\cite{11,18}. In other words, the machinery completed by Banerjee et al. can be utilized to extract information such as first passage probability density of ``correct''products among the coexistence of two types of ligands, which is different from our case. Basically, we follow the recipe from Polizzi et al for the calculation of the MFPT\cite{19}. The detailed procedure to obtain the passage time is given in the Appendix C. 

\begin{figure}[h] 
	\subfigure[]{
		\includegraphics[width=0.5\textwidth]{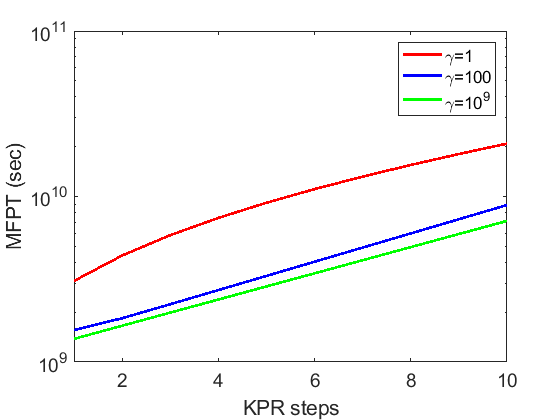}
	}
	\quad
	\subfigure[]{
		\includegraphics[width=0.5\textwidth]{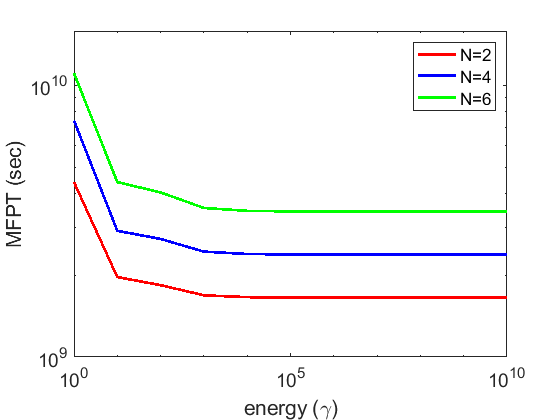}			
	}
	\caption{\textit{(a) The mean first passage time(MFPT) with variation of KPR steps measured at different energy $\gamma$=1, 100, and 10$^9$ respectively. (b) The MPFT decreases with energy input.}}
\end{figure}

\indent As shown in Figure 5(a) and (b), the numerical results reveal that the MFPT increases with the formation of more phosphorylated complexes as expected, but decreases with the energy input. There exist two phases characterized by a steep drop of the escape time until $\gamma$ reaches 10 and a modest drop in the regime of energy(10$<\gamma<10^{4}$) for any KPR step. After the point, the MFPT remains almost constant.\par
\indent The lowest value of the escape time in high energy regime where the lowest error fractions are also achieved manifestly conflicts with a general belief in a trade-off between accuracy and speed. There are a few studies on elucidating a broken compromise between accuracy and speed\cite{10,11}. They characterize the optimal condition where both the lowest error rate and MFPT are attained under the change in a certain kinetic parameter.  However, in a different context, we want to discuss such a trade-off in terms of energy input in a qualitative manner. Our interest is to investigate the role of compromise between accuracy and speed when energy input is taken into consideration with variations of several parameters. However, it may not be possible to characterize all the conditions for which a broken compromise holds due to numerous sets of different values of kinetic parameters. As indicated from Figure 2(b) and 5(b), one reveals the break down of the trade-off between error fractions and MFPT. However, when the dissociation rate constant increases from 0.1 to 0.5, we observe the passage time grows and reaches a converged value with energy, which reads 5.80x$10^{9}$ and 9.28x$10^{10}$ secs respectively for N=2 and N=6 cases, which is in contrast to the trend featuring its decrease with energy subject to the parameter we originally present. Another study by decreasing the forward rate constant from 0.5 to 0.1 gives a similar result. A study on T cell activation shows that the estimated MFPT which varies with energy consumption is approximately $10^{5}$ to $10^{13}$ secs\cite{10}. 

\vspace{5mm}

\section{Result: Adaptation Errors}

\indent In biological systems, a stimulus signal generates corresponding outcomes. The change in output in response to the perturbation allows the systems to return to the original one whose output is measured without a signal input\cite{20}. For T cell recognition, a sudden shift of a given parameter leads to change in an output activity to some extent despite its eventual recovery. It is meaningful to find out how accurately a perturbed system returns to the unperturbed one varying with KPR steps and energy consumption.

\begin{figure}[!] 
	\subfigure[]{
		\includegraphics[width=0.5\textwidth]{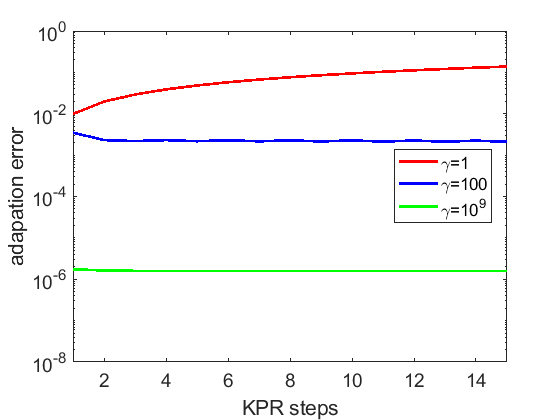}
	}
	\quad
	\subfigure[]{
		\includegraphics[width=0.5\textwidth]{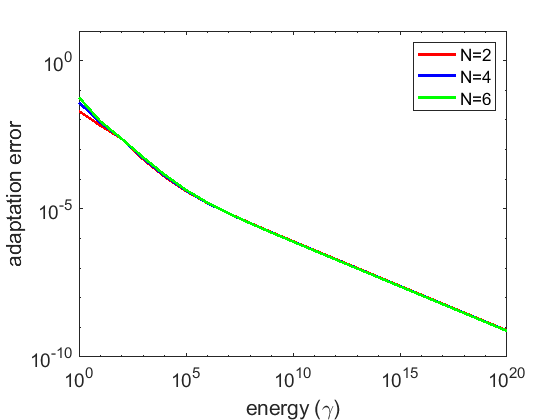}			
	}
	\caption{\textit{(a) The adaptation errors displays their virtual independence from the KPR steps when energy is applied. However, under the detailed balance condition($\gamma$=1), we observe its slight increase with KPR steps. (b) Energy utilization is highly responsible for the significant drop of the adaptation error.}}
\end{figure}

\indent We take an ``adaptation error'' in order to quantify such a biological adaptation which is defined as $\mid \frac{a_{0}-a}{a_{0}}\mid$, where $a_{0}$ being the amount of change in output activity without perturbation and a being the amount of change in output activity due to perturbation, and the error is expected to decline with energy cost\cite{20}. Among a few candidates as an input signal leading to a decline of the adaptation error, we have found that a slight change in the backward rate is highly responsible for reducing the adaptation error(See Figure 6(b)). \\
\indent The initial perturbation by either growth or drop of the backward rate changes the output activity that reads the final concentration of all proteins accordingly, but the energy input regulates the variation of the concentration, leading to a recovery of its original value. We also compare the adaptation errors for the foreign and self cases. As shown in Figure 7, our numerical results show that there is a tiny difference of the adaptation errors varying with KPR steps between the two ligands under the detailed balance condition($\gamma$=1). However, it is found that there is a slight increase in the adaptation error with KPR steps for the self proteins when energy is consumed for $\gamma$=100, while the result shows the opposite consequence for the foreign antigens. The difference between the results of the adaptation error in terms of KPR steps for the two distinct ligands is remarkable in high energy regime. This implies that the T cell system is more susceptible to the exposure to an attack of the self proteins in that the adaptation error is rather enhanced with KPR steps, which is in contrast to the results when only the foreign ligands are in the presence. We also find 10-fold increment of the backward rate as the perturbation increases the adaptation errors by approximately 10 times in the whole range of KPR steps and energy. In addition to this, it is observed that a slight change in the kinetic parameters is not a critical factor that influences a general trend of the adaptation error. For example, when the forward rate doubles, the error decreases by only 18.8\%.

\begin{figure}[h] 
	\centering	
	\includegraphics[width=0.5\textwidth]{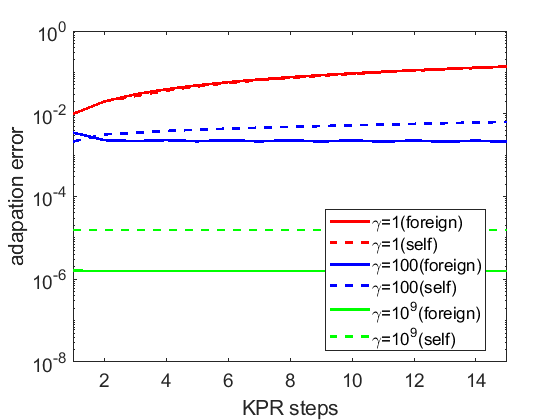}	
	\caption{\textit{Comparison of adaptation error between foreign and self ligands varying with KPR steps at different energy levels, $\gamma$=1, 100, and $10^9$}}	
\end{figure}

\vspace{30mm}

\section{Discussion}

\indent  It is difficult to predict the consequences of T cell dynamics without numerical calculation due to the complexity of our T cell scheme. For example, the dissociation event at each intermediate product and the direct process forming a phosphorylated complex without passing through previous intermediate stage are necessary elements to understand T cell recognition, as well as forward and backward rates between two products. Moreover, considering the nonequilibrium nature of living organisms, interacting with environments constantly, we had to incorporate energy source associated with ATP hydrolysis into our system. As used by Qian\cite{9}, the utilized energy is expressed in terms of several kinetic rate constants, and it indicates that most of the rates governing our T cell system depend on the consumed energy, which makes the related dynamics more complex. Hence, it is important to take all the information into account to set up an appropriate model for understanding T cell recognition. \\
\indent Despite existing studies on kinetic proofreading in T cell recognition, the lack of simultaneous comparisons of physical outcomes has prevented us from fully understanding the dynamics of the process in terms of energy input. As part of addressing such a problem, we present all the results regarding error rates, sensitivity, specificity, speed and adaptation errors in terms of energy cost with given KPR steps. \\
\indent It has been found that the error fractions decrease with energy utilization and KPR steps, and they have asymptotic behaviors, converging to certain minimum values when a sufficient amount of energy is supplied. Compared to the numerical results of specificity, we also have found that the error rates determined at certain amount of energy consumption lead to maximized specificity. 
In addition to this, the energy supply plays a critical role in reducing the escape time, accelerating the speed of signal transduction by minimizing the time-delay caused by the growth of KPR steps under the given kinetic scheme. In other words, we find a condition where a trade-off between error rates and MFPT is broken. Finally, we find that when backward rate is perturbed, our T cell system is fully adapted, which is characterized by computing the adaptation errors. \\
\indent Our kinetic model for T cell recognition with given set of parameters features an optimal condition where the lowest error fractions and the highest specificity with the fastest speed when certain amount of energy supplied. 

\vspace{5mm}

\begin{figure*}[h] 
	\centering     
	\tikzset{main node/.style={circle,draw,font=\sffamily\Large\bfseries}}  
	\tikzset{text node/.style={inner sep=2pt,font=\sffamily\Large\bfseries}}  
	\begin{tikzpicture}[->,>=stealth', shorten >=1pt, auto,
	node distance=2.5cm, thick, scale=0.75, transform shape, align=center,
	state/.style={circle, draw, minimum size=1.5cm}]
	
	\node[text node,red] (1) {TCR+Ligand};
	\node[state,main node,blue] (2) [right of=1,node distance=4cm] {$C_{0}$};
	\node[state,main node,blue] (3) [right of=2] {$C_{1}$};
	\node[state,main node,blue] (4) [right of=3] {$C_{2}$};
	\node[state,main node,blue] (5) [right of=4] {...};
	\node[state,main node,blue] (6) [right of=5] {$C_{N-1}$};
	\node[state,main node,blue] (7) [right of=6] {$C_{N}$};
	\node[state,main node,red] (8) [right of=7] {$absorb$};

	\def\myshift{2pt}
	
	\path[every node/.style={font=\sffamily\small}]
	([yshift=\myshift]1.0) edge node [above] {$k_{1}$} ([yshift=\myshift]2.180)
	([yshift=\myshift]2.0) edge node [above] {$k_{p}$} ([yshift=\myshift]3.180)
	([yshift=\myshift]3.0) edge node [above] {$k_{p}$} ([yshift=\myshift]4.180)
	([yshift=\myshift]4.0) edge node [above] {$...$}   ([yshift=\myshift]5.180)
	([yshift=\myshift]5.0) edge node [above] {$...$}   ([yshift=\myshift]6.180)
	([yshift=\myshift]6.0) edge node [above] {$k_{p}$} ([yshift=\myshift]7.180)
	([yshift=\myshift]7.0) edge node [above] {$W$}     ([yshift=\myshift]8.180)

	([yshift=-\myshift]2.180) edge node [below] {$k_{-1}$} ([yshift=-\myshift]1.0)
	([yshift=-\myshift]3.180) edge node [below] {$k_{-p}$} ([yshift=-\myshift]2.0)
	([yshift=-\myshift]4.180) edge node [below] {$k_{-p}$} ([yshift=-\myshift]3.0)  
	([yshift=-\myshift]5.180) edge node [below] {...}    ([yshift=-\myshift]4.0)  
	([yshift=-\myshift]6.180) edge node [below] {...}    ([yshift=-\myshift]5.0) 
	([yshift=-\myshift]7.180) edge node [below] {$k_{-p}$} ([yshift=-\myshift]6.0)

	(3) edge [bend right=45,pos=0.3] node [above] {$k_{disso}$}   (1.100)
	(4) edge [bend right=45,pos=0.3] node [above] {$k_{disso}$}   (1.135)
	(5) edge [bend right=45,pos=0.3] node [above] {...}   (1.150)
	(6) edge [bend right=45,pos=0.3] node [above] {$k_{disso}$} (1.158)
	(7) edge [bend right=45,pos=0.3] node [above] {$k_{disso}$}  (1.165)

	(1) edge [bend right=45,pos=0.5] node [below] {$m_{1}$} (3)
	(1) edge [bend right=45,pos=0.5] node [below] {$m_{2}$} (4)
	(1) edge [bend right=45,pos=0.3] node [above] {...} (5)
	(1) edge [bend right=45,pos=0.5] node [below] {$m_{N-1}$} (6)
	(1) edge [bend right=45,pos=0.5] node [below] {$m_{N}$} (7);
	
	\end{tikzpicture}	
\end{figure*}

\FloatBarrier

\section{Appendix A}

We apply the mass action law to express the time derivative of concentration for each bound state, which is given by

\begin{equation}
\begin{aligned}
\frac{dC_{0}}{dt}&=k_{1}[R]-(k_{-1}+k_{p})C_{0}+k_{-p}(\gamma)C_{1} \\
\frac{dC_{1}}{dt}&=k_{p}C_{0}-(k_{-p}(\gamma)+k_{p}+k_{disso})C_{1}+k_{-p}(\gamma)C_{2}+\frac{k_{p}k_{1}k_{disso}}{\gamma k_{-p}(\gamma)k_{-1}}[R] \\
\frac{dC_{2}}{dt}&=k_{p}C_{1}-(k_{-p}(\gamma)+k_{p}+k_{disso})C_{2}+k_{-p}(\gamma)C_{3}+\frac{k_{p}m_{1}}{\gamma k_{-p}(\gamma)}[R] \\
... \\
\frac{dC_{N-1}}{dt}&=k_{p}C_{N-2}-(k_{-p}(\gamma)+k_{p}+k_{disso})C_{N-1}+k_{-p}(\gamma)C_{N} +\frac{k_{p}m_{N-2}}{\gamma k_{-p}(\gamma)}[R] \\
\frac{dC_{N}}{dt}&=k_{p}C_{N-1}-(k_{-p}(\gamma)+k_{disso}+W)C_{N}+\frac{k_{p}m_{N-1}}{\gamma k_{-p}(\gamma)}[R]  
\end{aligned}
\end{equation} 


Here, [R] denotes the concentration of unbound TCR. Again, the concentration of free ligands are absorbed into rate constants $k_{1}$ and $m_{i}$. \par
Applying the SSA to each intermediate including the final complex that contributes to signaling, we get

\begin{equation}
C_{0}=\frac{k_{1}[R]+k_{-p}(\gamma)C_{1}}{k_{-1}+k_{p}} 
\end{equation}

\begin{equation}
C_{1}=\frac{k_{p}C_{0}+k_{-p}(\gamma)C_{2}+\frac{k_{p}k_{1}k_{disso}}{\gamma k_{-p}(\gamma)k_{-1}}[R]}{k_{-p}(\gamma)+k_{p}+k_{disso}} 
\end{equation}

\begin{equation}
C_{2}=\frac{k_{p}C_{1}+k_{-p}(\gamma)C_{3}+\frac{k_{p}m_{1}}{\gamma k_{-p}(\gamma)}[R]}{k_{-p}(\gamma)+k_{p}+k_{disso}} 
\end{equation}

The general expression for $C_{N-1}$ just before the formation of a final complex is as follows.
\begin{equation}
C_{N-1}=\frac{k_{p}C_{N-2}+k_{-p}(\gamma)C_{N}+\frac{k_{p}m_{N-2}}{\gamma k_{-p}}[R]}{k_{-p}(\gamma)+k_{p}+k_{disso}}
\end{equation}

The initial concentration $C_{0}$ given by above can be replaced by $C_{0}=\frac{k_{1}[R]}{k_{-1}}$ assuming $k_{-1}>>k_{p}$ and $k_{1}[R]>>k_{-p}C_{1}$\cite{9}. \\

The concentration at the final state is given by
\begin{equation}
C_{N}=\frac{k_{p}C_{0}+\frac{k_{p}k_{1}k_{disso}}{\gamma k_{-p}(\gamma)k_{-1}}[R]}{k_{-p}(\gamma)+k_{disso}+W}  \quad  (N=1);  
\end{equation}

\begin{equation}
\mbox{\normalsize$C_{N}=\frac{k_{p}^{2}C_{N-2}+k_{p}\frac{k_{p}m_{N-2}}{\gamma k_{-p}(\gamma)}[R]+\frac{k_{p}m_{N-1}}{\gamma k_{-p}(\gamma)}[R]\left(k_{-p}(\gamma)+k_{p}+k_{disso}\right)}{\left(k_{-p}(\gamma)+k_{disso}+W\right)\left(k_{-p}+k_{p}+k_{disso}\right)-k_{p}k_{-p}(\gamma)}$}   \quad  (N>1)  
\end{equation} 

\indent Note that each series of $C_{N}$ depends on the number of KPR steps, whose expression for N$>$2 case has a recursion relationship that connects with a $C_{N-2}$ term, generating additional terms successively (i.e.) $C_{N-4}, C_{N-6}$, and so on ending with $C_{0}$ for n=even and $C_{1}$ for n=odd. \par
On the other hand, $C_{N-2}=\frac{k_{p}C_{N-3}+k_{-p}(\gamma)C_{N-1}+\frac{k_{p}m_{N-3}}{\gamma k_{-p}(\gamma)}[R]}{k_{-p}(\gamma)+k_{p}+k_{disso}}$ with substitution of the expression for $C_{N-1}$ and $C_{N-3}$ respectively. Solving for $C_{N-2}$, we get

\begin{strip}
\begin{equation}
\begin{aligned}
\mbox{\Large$C_{N-2}=\frac{k_{p}}{k_{-p}(\gamma)+k_{p}+k_{disso}}\left[\frac{k_{p}C_{N-4}+k_{-p}C_{N-2}+\frac{k_{p}m_{N-4}[R]}
{\gamma k_{-p}(\gamma)}}{k_{-p}(\gamma)+k_{p}+k_{disso}}\right]+\frac{k_{-p}(\gamma)}{k_{-p}(\gamma)+k_{p}+k_{disso}}\left[\frac{k_{p}C_{N-2}+k_{-p}(\gamma)C_{N}+m_{N-1}[R]}{k_{-p}(\gamma)+k_{p}+k_{disso}}\right]+\frac{\frac{k_{p}m_{N-3}}{\gamma k_{-p}}[R]}{k_{-p}(\gamma)+k_{p}+k_{disso}}$}
\end{aligned}
\end{equation}
\end{strip}



\indent This becomes, \\




\begin{strip}
\begin{equation}
\begin{aligned}
\mbox{\Large$C_{N-2}=\frac{k_p^2C_{N-4}+\left(k_{p}m_{N-3}+k_{-p}(\gamma)m_{N-1}\right)[R]+\frac{k_{p}k_{-p}(\gamma)^2 \frac{k_{p}m_{N-2}}{\gamma k_{-p}(\gamma)}[R]+k_{-p}(\gamma)^2 \frac{k_{p}m_{N-1}}{\gamma k_{-p}}[R]\left(k_{-p}(\gamma)+k_{p}+k_{disso}\right)}{\left(k_{-p}(\gamma)+k_{disso}+W)(k_{-p}(\gamma)+k_{p}+k_{disso}\right)-k_{p}k_{-p}(\gamma)}}{\left(k_{-p}(\gamma)+k_{p}+k_{disso}\right)^{2}-2k_{p}k_{-p}(\gamma)-\frac{k_{p}^2 k_{-p}(\gamma)^2}{\left(k_{-p}(\gamma)+k_{disso}+W\right)\left(k_{-p}(\gamma)+k_{p}+k_{disso}\right)-k_{p}k_{-p}(\gamma)}}$} \\
\mbox{\normalsize$+\frac{\frac{k_{p}m_{N-3}}{\gamma k_{-p}(\gamma)}[R]\left(k_{-p}(\gamma)+k_{p}+k_{disso}\right)} 
{\left(k_{-p}(\gamma)+k_{p}+k_{disso}\right)^{2}-2k_{p}k_{-p}(\gamma)-\frac{k_{p}^2k_{-p}(\gamma)^2}{\left(k_{-p}(\gamma)+k_{disso}+W\right)\left(k_{-p}(\gamma)+k_{p}+k_{disso}\right)-k_{p}k_{-p}(\gamma)}}$}
\end{aligned}
\end{equation}  
\end{strip}

The initial concentrations generated by the above recursion relation are expressed as follows. \par

\mbox{}

\begin{strip}
\begin{equation}
\begin{aligned}
\mbox{\large$C_{1}=\frac{\left[(k_{-p}(\gamma)+k_{p}+k_{disso})k_{p}(k_{1}/k_{-1})+k_{-p}(\gamma)\frac{k_{p}m_{1}}{\gamma k_{-p}}\right][R]+\frac{k_{p}k_{-p}(\gamma)^2 \frac{k_{p}m_{1}}{\gamma k_{-p}}[R]+k_{-p}(\gamma)^2 \frac{k_{p}m_{2}}{\gamma k_{-p}}[R]\left(k_{-p}(\gamma)+k_{p}+k_{disso}\right)}{\left(k_{-p}(\gamma)+k_{disso}+W\right)\left(k_{-p}(\gamma)+k_{p}+k_{disso}\right)-k_{p}k_{-p}(\gamma)}}{\left(k_{-p}(\gamma)+k_{p}+k_{disso}\right)^{2}-k_{p}k_{-p}(\gamma)-\frac{k_{p}^2 k_{-p}(\gamma)^2}{\left(k_{-p}(\gamma)+k_{disso}+W\right)\left(k_{-p}(\gamma)+k_{p}+k_{disso}\right)-k_{p}k_{-p}(\gamma)}}$} \\
\mbox{\Large$+\frac{m_{1}[R]\left(k_{-p}(\gamma)+k_{p}+k_{disso}\right)}{\left(k_{-p}(\gamma)+k_{p}+k_{disso}\right)^{2}-k_{p}k_{-p}(\gamma)-\frac{k_{p}^2 k_{-p}(\gamma)^2}{\left(k_{-p}(\gamma)+k_{disso}+W\right)\left(k_{-p}(\gamma)+k_{p}+k_{disso}\right)-k_{p}k_{-p}(\gamma)}}$}  \quad (N=odd)
\end{aligned}
\end{equation}
\end{strip}

\begin{equation}
\mbox{\large$C_{0}=\frac{k_{1}[R]}{k_{-1}}$} \quad (N=even)
\end{equation}

The error fraction f is defined as the ratio of the rate of ``wrong'' product formation to the rate of ``correct'' product formation (i.e.) $\frac{C_{N,self}}{C_{N,foreign}}$ for T-Cell targeting\cite{1,2,16}. Therefore, the full expression of error fraction for our N-cycle kinetic proofreading model is given by \par

\begin{equation}
\mbox{\large$f=\frac{\left(k_{p}C_{0,self}+\frac{k_{p}k_{1}k_{disso}/\theta}{\gamma k_{-p}(\gamma) k_{-1}}[R]\right)\left(k_{-p}(\gamma)+k_{disso}+W\right)}{\left(k_{p}C_{0,foreign}+\frac{k_{p}k_{1}k_{disso}}{\gamma k_{-p}(\gamma)k_{-1}}[R]\right)\left(k_{-p}(\gamma)+\frac{k_{disso}}{\theta}+W\right)}$}   \quad (N=1)
\end{equation}

\begin{equation}
\mbox{\large$f=\frac{\frac{k_{p}^{2}C_{N-2,self}+k_{p}\frac{k_{p}m_{N-2}}{\gamma k_{-p}(\gamma)}[R]+\frac{k_{p}m_{N-1}}{\gamma k_{-p}(\gamma)}[R]\left(k_{-p}(\gamma)+k_{p}+\frac{k_{disso}}{\theta}\right)}{\left(k_{-p}(\gamma)+\frac{k_{disso}}{\theta}+W\right)\left(k_{-p}(\gamma)+k_{p}+\frac{k_{disso}}{\theta}\right)-k_{p}k_{-p}(\gamma)}}{\frac{k_{p}^{2}C_{N-2,foreign}+k_{p}\frac{k_{p}m_{N-2}}{\gamma k_{-p}(\gamma)}[R]+\frac{k_{p}m_{N-1}}{\gamma k_{-p}(\gamma)}[R]\left(k_{-p}(\gamma)+k_{p}+k_{disso}\right)}	{\left(k_{-p}(\gamma)+k_{disso}+W\right)\left(k_{-p}(\gamma)+k_{p}+k_{disso}\right)-k_{p}k_{-p}(\gamma)}}$}  \quad (N>1)
\end{equation}

, where \\




\begin{strip}
\begin{equation}
\begin{aligned}
\mbox{\large$C_{N-2,foreign}=\frac{k_p^2C_{N-4}+\left(k_{p}\frac{k_{p}m_{N-4}}{\gamma k_{-p}(\gamma)}+k_{-p}(\gamma)\frac{k_{p}m_{N-2}}{\gamma k_{-p}(\gamma)}\right)[R]+\frac{k_{p}k_{-p}(\gamma)^2\frac{k_{p}m_{N-2}}{\gamma k_{-p}(\gamma)}[R]+k_{-p}(\gamma)^2\frac{k_{p}m_{N-1}}{\gamma k_{-p}(\gamma)}[R]\left(k_{-p}(\gamma)+k_{p}+k_{disso}\right)}{\left(k_{-p}(\gamma)+k_{disso}+W\right)\left(k_{-p}(\gamma)+k_{p}+k_{disso}\right)-k_{p}k_{-p}(\gamma)}} 
{\left(k_{-p}(\gamma)+k_{p}+k_{disso}\right)^{2}-2k_{p}k_{-p}(\gamma)-\frac{k_{p}^2k_{-p}(\gamma)^2}{\left(k_{-p}(\gamma)+k_{disso}+W\right)\left(k_{-p}(\gamma)+k_{p}+k_{disso}\right)-k_{p}k_{-p}(\gamma)}}$} \\
\mbox{\Large$+\frac{\frac{k_{p}m_{N-3}}{\gamma k_{-p}(\gamma)}[R]\left(k_{-p}(\gamma)+k_{p}+k_{disso}\right)}	
{\left(k_{-p}(\gamma)+k_{p}+k_{disso}\right)^{2}-2k_{p}k_{-p}(\gamma)-\frac{k_{p}^2k_{-p}(\gamma)^2}{\left(k_{-p}(\gamma)+k_{disso}+W\right)\left(k_{-p}(\gamma)+k_{p}+k_{disso}\right)-k_{p}k_{-p}(\gamma)}}$}
\end{aligned}
\end{equation} 
\end{strip} 

\mbox{}

\begin{strip}
\begin{equation}
\begin{aligned}
\mbox{\large$C_{N-2,self}=\frac{k_p^2C_{N-4}+\left(k_{p}\frac{k_{p}m_{N-4}}{\gamma k_{-p}(\gamma)}+k_{-p}(\gamma)\frac{k_{p}m_{N-2}}{\gamma k_{-p}(\gamma)}\right)[R]+\frac{k_{p}k_{-p}(\gamma)^2\frac{k_{p}m_{N-2}}{\gamma k_{-p}(\gamma)}[R]+k_{-p}(\gamma)^2\frac{k_{p}m_{N-1}}{\gamma k_{-p}(\gamma)}[R]\left(k_{-p}(\gamma)+k_{p}+\frac{k_{disso}}{\theta}\right)}{\left(k_{-p}(\gamma)+\frac{k_{disso}}{\theta}+W\right)\left(k_{-p}(\gamma)+k_{p}+\frac{k_{disso}}{\theta}\right)-k_{p}k_{-p}(\gamma)}}{\left(k_{-p}(\gamma)+k_{p}+\frac{k_{disso}}{\theta}\right)^{2}-2k_{p}k_{-p}(\gamma)-\frac{k_{p}^2k_{-p}(\gamma)^2}{\left(k_{-p}(\gamma)+\frac{k_{disso}}{\theta}+W\right)\left(k_{-p}(\gamma)+k_{p}+\frac{k_{disso}}{\theta}\right)-k_{p}k_{-p}(\gamma)}}$} \\
\mbox{\Large$+\frac{m_{N-2}[R]\left(k_{-p}(\gamma)+k_{p}+\frac{k_{disso}}{\theta}\right)}	
{\left(k_{-p}(\gamma)+k_{p}+\frac{k_{disso}}{\theta}\right)^{2}-2k_{p}k_{-p}(\gamma)-\frac{k_{p}^2k_{-p}(\gamma)^2}{\left(k_{-p}(\gamma)+\frac{k_{disso}}{\theta}+W\right)\left(k_{-p}(\gamma)+k_{p}+\frac{k_{disso}}{\theta}\right)-k_{p}k_{-p}(\gamma)}}$}
\end{aligned}
\end{equation}  
\end{strip}

Again, the expression for $C_{N-2}$ can be given in terms of either $C_{0}$ for even n or $C_{1}$ for odd n.

\vspace{5mm}

\section{Appendix B}

We present the analytical expression for both the sensitivity and the specificity for our kinetic model in terms of the energy($\gamma$). The general forms for sensitivity and specificity which are expressed in the main text is given by the following. \\

\begin{equation}
\textnormal{Sensitivity}=\frac{C_{total,foreign}\;\alpha_{correct}^{N}}{C_{total,foreign}\;\alpha_{correct}^{N}+C_{total,foreign}\;\left(1-\alpha_{correct}^{N}\right)}
\end{equation}

\begin{equation}
\textnormal{Specificity}=\frac{C_{total,foreign}\;\alpha_{correct}^{N}}{C_{total,foreign}\;\alpha_{correct}^{N}+C_{total,self}\;\alpha_{wrong}^{N}}
\end{equation}

, which can be reduced to the simple forms. \\

\begin{equation}
\textnormal{Sensitivity} = \alpha_{foreign}^{N} = \frac{C_{N,foreign}}{C_{total,foreign}} 
\end{equation}

with 

\begin{equation}
C_{total,foreign} = C_{0,foreign} +  C_{N,foreign} + \sum_{i=1}^{N-1} C_{i,foreign} 
\end{equation}

\vspace{10mm}

\begin{equation}
\textnormal{Specificity} = \frac{C_{N,foreign}}{C_{N,foreign}+C_{N,self}}
\end{equation}

, where $C_{i, foreign}$ (i$<$N) is the concentration of each foreign intermediate complex except for the final foreign product which is associated with signaling. $C_{N}$ is the concentration of the final complex, which is given by \\

\begin{equation}
C_{N}=\frac{k_{p}C_{0}+\frac{k_{p}k_{1}k_{disso}}{\gamma k_{-p}(\gamma)k_{-1}}[R]}{k_{-p}(\gamma)+k_{disso}+W}  \quad  (N=1);  
\end{equation}

\begin{equation}
\mbox{\large$C_{N}=\frac{k_{p}^{2}C_{N-2}+k_{p}\frac{k_{p}m_{N-2}}{\gamma k_{-p}(\gamma)}[R]+\frac{k_{p}m_{N-1}}{\gamma k_{-p}(\gamma)}[R]\left(k_{-p}(\gamma)+k_{p}+k_{disso}\right)}{\left(k_{-p}(\gamma)+k_{disso}+W\right)\left(k_{-p}+k_{p}+k_{disso}\right)-k_{p}k_{-p}(\gamma)}$}   \quad  (N>1)  
\end{equation}

\vspace{20mm}

\section{Appendix C} 

All states including the initial states characterized by free receptors and ligands and all intermediate complex are taken into consideration in order to calculate the MFPT. The noteworthy consequence of the MFPT calculation done by Polizzi et al. elucidates that the MFPT is the sum of the residence of each state\cite{19}. We follows the main steps from their work. 
We define the probability density $f_{i}(t)$ as follows: \\
\begin{equation}
f_{i}(t) = \frac{d}{dt} P_{pass}(t)
\end{equation}

Also, noting that the probability of making a first passage to the ``absorb'' state at time t is given by 1-(sum of probability of making a first passage to each state). in other words,
\begin{equation}
P_{pass}= 1-\sum_{j\neq absorb}P_{j}(t)
\end{equation}

We express the probability density in terms of the $P_{pass}$ defined above. Then, the MFPT expression is \\
\begin{equation}
<\tau>=\int_{0}^{\infty} t f(t) dt = -\int_{0}^{\infty} \sum_{j\neq absorb} t \,\frac{d}{dt} P_{j} (t) dt 
\end{equation}

The boundary term vanishes after performing the integration by parts. This becomes,

\begin{equation}
<\tau>=\sum_{i \neq absorb}\int_{0}^{\infty} P_{i}(t) dt = \sum_{i \neq absorb} r_{i}
\end{equation}

Since the first order kinetics controls the entire system, we get

\begin{equation}
P_{i} = \left[exp(Kt) P_{0}\right]_{i}
\end{equation}

\noindent Here, K denotes the rate matrix and P and ${P_{0}}$ represent the state population vector and the initial condition vector respectively.

\noindent This indicates,
\begin{equation}
r_{i}=\left[\int_{0}^{\infty} exp(Kt) P_{0} dt\right]_{i} = \left[-K^{-1} P_{0}\right]_{i}
\end{equation}

The governing equation expressed as $\displaystyle{\frac{d}{dt} \textnormal{C(t)} =\textnormal{K C(t)}}$ gives the series of the initial concentration denoted by $C_{T}$ , $C_{0}$, $C_{1}$ ... $C_{N}$, $C_{absorb}$ for our T cell model where K is

\begin{strip}
\begin{equation}
\textbf{K}=
\newcommand\scalemath[2]{\scalebox{#1}{\mbox{\ensuremath{\displaystyle #2}}}}
\scalemath{0.88}{\begin{pmatrix}
	-\left(k_{1}+\sum\limits_{i=1}^{N} m_{i}\right) & k_{-1} & k_{disso} & ... & k_{disso} & k_{disso} & 0 \\
	k_{1} & -(k_{-1}+k_{p})  & k_{-p}(\gamma)  & ... & 0 & 0 & 0  \\
	\frac{k_{p}k_{1}k_{disso}}{\gamma k_{-p}(\gamma)k_{-1}} & k_{p} & -(k_{-p}(\gamma)+k_{p}+k_{disso}) & ... & 0 & 0 & 0 \\
	\frac{k_{p}m_{1}}{\gamma k_{-p}(\gamma)} & 0 & k_{p} & ... & 0 & 0 & 0 \\
	... \\
	\frac{k_{p}m_{N-2}}{\gamma k_{-p}(\gamma)} & 0 & 0 & ... & -(k_{-p}(\gamma)+k_{p}+k_{disso}) & k_{-p}(\gamma) & 0  \\
	\frac{k_{p}m_{N-1}}{\gamma k_{-p}(\gamma)} & 0 & 0 & ... & k_{p} & -(k_{-p}(\gamma)+k_{disso}+W) & 0 \\
	0 & 0 & 0 & ... & 0 & W & 0  
	\end{pmatrix}}
\end{equation}
\end{strip}

A Matlab software is used in order to compute the MFPT by summing over the product of $-K^{-1}C_{T}$ defined at each state except for ``absorb''. For example, $r_{i}= \left[-K^{-1}C_{T}\right]_{i}$ for N=1 case is given by

\begin{equation}
\setstackgap{L}{1.8\baselineskip}
\mathbf{r_{1}} = C_{T}\bracketVectorstack{\frac{\left(k_{p}+k_{-1}\right)W+\left(k_{p}+k_{-1}\right)k_{disso}+k_{-1}k_{-p}}{\left(k_{1}k_{p}+m_{1}k_{p}+m_{1}k_{-1}\right)W} \\
	\frac{k_{1}W+k_{1}k_{disso}+k_{1}k_{-p}+m_{1}k_{-p}}{\left(k_{1}k_{p}+m_{1}k_{p}+m_{1}k_{-1}\right)W} \\
	\frac{1}{W}}
\end{equation}

\vspace{5mm}

\section{Appendix D}  

The adaptation error we define for our kinetic model is given by the following. \\

\begin{equation}
\textnormal{Adp error} = \left|\frac{C_{final, perturbed}-C_{final, unperurbed}}{C_{final, unperturbed}}\right|
\end{equation}

Here, $C_{final}$ is the concentration of the combined final products (foreign and self) whose expression is given by \par

\begin{equation}
\begin{aligned}
\mbox{\Large$C_{N}=\frac{k_{p}C_{0,foreign}+\frac{k_{p}k_{1}k_{disso}}{\gamma k_{-p}(\gamma)k_{-1}}[R]}{k_{-p}(\gamma)+k_{disso}+W} 
+ \frac{k_{p}C_{0,self}+\frac{k_{p}k_{1}k_{disso}/\theta}{\gamma k_{-p}(\gamma)k_{-1}}[R]}{k_{-p}(\gamma)+k_{disso}/\theta+W}$} \\
\quad  (N=1)  
\end{aligned}
\end{equation}

\begin{equation}
\begin{aligned}
\mbox{\Large$C_{N}=\frac{k_{p}^{2}C_{N-2,foreign}+k_{p}\frac{k_{p}m_{N-2}}{\gamma k_{-p}(\gamma)}[R]+\frac{k_{p}m_{N-1}}{\gamma k_{-p}(\gamma)}[R]\left(k_{-p}(\gamma)+k_{p}+k_{disso}\right)}{\left(k_{-p}(\gamma)+k_{disso}+W\right)\left(k_{-p}(\gamma)+k_{p}+k_{disso}\right)-k_{p}k_{-p}(\gamma)}$} \\
+ \mbox{\Large$\frac{k_{p}^{2}C_{N-2,self}+k_{p}\frac{k_{p}m_{N-2}}{\gamma k_{-p}(\gamma)}[R]+\frac{k_{p}m_{N-1}}{\gamma k_{-p}(\gamma)}[R]\left(k_{-p}(\gamma)+k_{p}+k_{disso}/\theta\right)}{\left(k_{-p}(\gamma)+k_{disso}+W\right)\left(k_{-p}(\gamma)+k_{p}+k_{disso}/\theta\right)-k_{p}k_{-p}(\gamma)}$} \\  
\quad  (N>1)
\end{aligned}
\end{equation}

$C_{final, perturbed}$ is the concentration which is determined at the slight changed value of the $k_{-p}$. 

\vspace{10mm} 

\section*{Acknowledgements}

\indent{\normalsize We thank supports from National Science Foundation(NSF-Phys.76066 and NSF-CHE-1808474).}  

\vspace{10mm}
\nocite{1,2,3,4,5,6,7,8,9,10,11,12,13,14,15,16,17,18,19,20}
\bibliography{PCCP_Ref}
\bibliographystyle{rsc}	
\end{document}